\documentclass[Afour,sageh,times,doublespace%
 reprint,
 amsmath,amssymb,
 aps,
]{sagej}

\usepackage{fontawesome}
\usepackage[dvipsnames]{xcolor}
\usepackage{moreverb,url}
\usepackage{graphicx}
\usepackage{dcolumn}
\usepackage{multirow}
\usepackage{bm}

\usepackage[colorlinks,bookmarksopen,bookmarksnumbered,citecolor=red,urlcolor=red]{hyperref}

\newcommand{\PP}{\colorbox{NavyBlue}{\color{white}\sf PP} }
\newcommand{\PSOE}{\colorbox{red}{\color{white}\sf PSOE} }
\newcommand{\Podemos}{\colorbox{Orchid}{\color{white}\sf Podemos} }
\newcommand{\Vox}{\colorbox{OliveGreen}{\color{white}\sf Vox} }
\newcommand{\Ciudadanos}{\colorbox{orange}{\color{white}\sf Ciudadanos} }

\def\volumeyear{2020}

\begin{document}

\runninghead{Folgado and Sanz}

\title{Exploring the political pulse of a country using data science tools}

\author{Miguel G. Folgado \affilnum{1} and Veronica Sanz\affilnum{1,2}}

\affiliation{\affilnum{1}Instituto de F\'isica Corpuscular (IFIC), Universidad de Valencia-CSIC, E-46980, Valencia, Spain \\
\affilnum{2} Department of Physics and Astronomy, University of Sussex, Brighton BN1 9QH, UK}

\corrauth{Veronica Sanz,  Department of Physics and Astronomy, University of Sussex, 
Brighton BN1 9QH, UK \\  Instituto de F\'isica Corpuscular (IFIC), Universidad de Valencia-CSIC, E-46980, Valencia, Spain}

\email{v.sanz@sussex.ac.uk, veronica.sanz@uv.es}

\begin{abstract}
In this paper we illustrate the use of Data Science techniques to analyse complex human communication. In particular, we consider tweets from leaders of political parties as a dynamical proxy to political programmes and ideas. We also study the temporal evolution of their contents as a reaction to specific events. We analyse levels of positive and negative sentiment in the tweets using new tools adapted to social media.  We also train an  Artificial Intelligence to recognise the political affiliation of a tweet. The AI is able to predict the origin of the tweet with a precision in the range of 71-75\%, and the political leaning (left or right) with a precision of around 90\%. This study is meant to be viewed as a proof-of-concept of interdisciplinary nature, at the interface between Data Science and political analysis. 
\end{abstract}


\maketitle

\section{Introduction}

Since the advent of its democracy in 1975, the political landscape in Spain had been strongly bi-partisan. The two main parties  representing the right and left sides of the political spectrum ---in their current incarnations \PP  and \PSOE---have typically received 80 \% or more of the votes. The rest of the political representation was made of  minority parties,  mostly focused on regional interests.

During the world financial crisis, which hit Spain the hardest in 2012, unemployment rose to about 30\% of the workforce and to about 60\% in the case of workers under 25 years of age~\cite{INE}.  That same year, the government took steps to bail out banks filing for bankruptcy, a decision which, together with  new austerity measures on  Spain's welfare system and the numerous high-profile corruption cases coming into light, may be responsible for  the breaking of the {\it status-quo}. The Spanish political scene since then has seen the rise of anti-establishment movements, sometimes in the form of citizen's platforms which coalesced into  political parties. 
\begin{figure}[h!]
    \centering
    \includegraphics[width=0.5\textwidth]{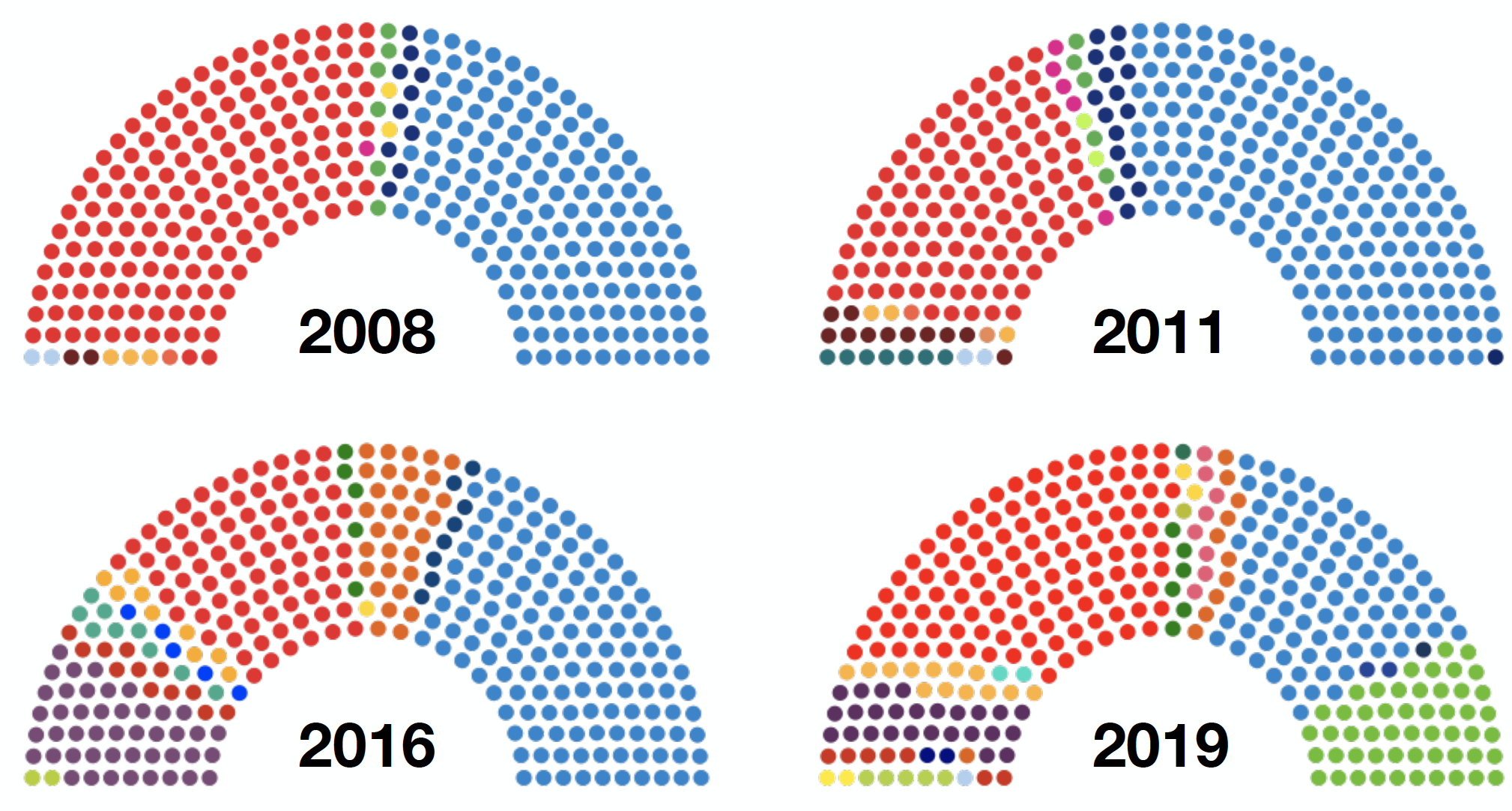}
    \caption{Four election results from the pre- and post-2012 period. The colours of the main five parties in current Spain do match the assignments in the text: \PSOE \PP \Vox \Podemos and \Ciudadanos. }
    \label{fig:evol}
\end{figure}

This evolution is reflected in the fragmentation of the  political spectrum shown in Fig.~\ref{fig:evol}. From 2012 on, the composition of the Congress of Deputies became much more {\it colourful}. Far-left \Podemos and far-right \Vox movements, as well as an alternative center party \Ciudadanos, emerged as leading minority groups in addition to traditional regional parties.  Generally speaking, post-2012 Spanish politics became more diverse, and more focused on identity issues. As regional tensions in the Spanish system ---already a strongly federal one---  were heightened, we witnessed the political growth of  Catalan and Basque separatist groups. 

In this study we show how modern data science can help characterizing the tenets represented by political parties, and their evolution. In other words, we  will use data to feel the {\it pulse} of Spanish politics. 
Beyond official party manifestos, the real pulse of politics expresses itself be found in media outlets such as Twitter. Indeed, we will show that one can use tweets from political leaders as a proxy for the parties' emotional state.   

We will start with a simple frequency analysis to extract each party's ideological bubble, then we move onto a more sophisticated analysis to quantify {\it sentiment} evolution, i.e. the level of negativity and positivity as a reaction to events. Finally, we show how,  using predictive AI tools, one can learn to identify party ideological adherence with the inspection of individual tweets.

\begin{table*}[t!]
\label{tab:tweetlist}
  \centering
  \begin{tabular}{lllll}
    \Podemos & \PSOE & \Ciudadanos & \PP & \Vox \\
    \hline
    \\
    @PabloIglesias & @sanchezcastejon & @InesArrimadas & @pablocasado\_ & @Santi\_ABASCAL \\
    @MiguelUrban & @CristinaNarbona & @MarinaBS\_Cs & @JavierMaroto & @Jorgebuxade \\
    @pilar\_lima & @Adrilastra & @CCuadradoCs & @ja\_nietob & @vicpiedra \\
    @anamarcelloana & @abalosmeco & @jmespejosaav & @AlmeidaPP\_ & @rromerovilches \\
    @pbustinduy & @santicl & @joanmesquida62 & @sanchezcesar & @Pablosaenzd \\
    @Alber\_Canarias & @JoseantonioJun & @MelisaRguezCs & @abeltran\_ana & @Vox\_Molina \\
    @ionebelarra & @patxilopez & @BalEdmundo & @javiermarquezsa & @MeerRocio \\
    @MayoralRafa & @oscar\_puente\_ & @Tonicanto1 & @vicentetiradopp & @Igarrigavaz \\
    @Julio\_Rodr\_ & @gomezdcelis & @Lroldansu & @DiegoCalvoPouso & @jlsteeg \\
    @VeraNoelia &  & @ignacioaguado & & @RuizSolas \\
    @IreneMontero &  & @Albert\_Rivera & & @pedro\_fhz \\
    @pnique &  & & &  \\
    @VicencNavarro &  & & &  \\
    \\
  \end{tabular}
  \caption{Twitter accounts selected for this study.}
  \label{tab:1}
\end{table*}

\section{Data selection and processing}\label{sec:dataselection}
Before we present our results, we describe the procedure to generate the datasets and the pre-processing needed to obtain quantitative results.

Our input data is the content of tweets. We have restricted ourselves to the five main~\footnote{Note that we are focusing on Spanish-wide parties. ERC, a separatist Catalonian party, did overtake \Ciudadanos in the November 2019 elections.} political parties: \Ciudadanos \Podemos \PP \PSOE and \Vox. Since we would like to focus on  the political situation since the bi-partism was broken, with new parties gaining national relevance, we examine tweets from January 2016.

To obtain the tweets we first enrolled as developers with Twitter (a relatively simple procedure for academic purposes) and then used standard tools to select and collect the tweets.  The most common tool is {\tt tweepy}~\cite{tweepy}, the official Twitter API which enables us to obtain tweets from different accounts.
There is an issue with this package, though, as it only allows the collection of the last 3200 tweets per account. Maybe this quantity would be huge for the most part of the population, but not for politicians, who display a constant presence in Twitter. To collect a larger tweet datasample  we used  another tool called {\tt GetOldTweets3}~\cite{getoldtweets}.

In order to avoid  subjective bias in the selection of candidates, we explore the main webs of the  political parties and examine the list of members of the internal committees. To obtain similar numbers of tweets for each party, we are forced to select different numbers of members in each case, depending on the activity of the different members of the committees, see Table 1. In particular, for \Podemos we select a larger number of representatives.

\begin{figure}[h!]
    \centering
    \includegraphics[width=0.5\textwidth]{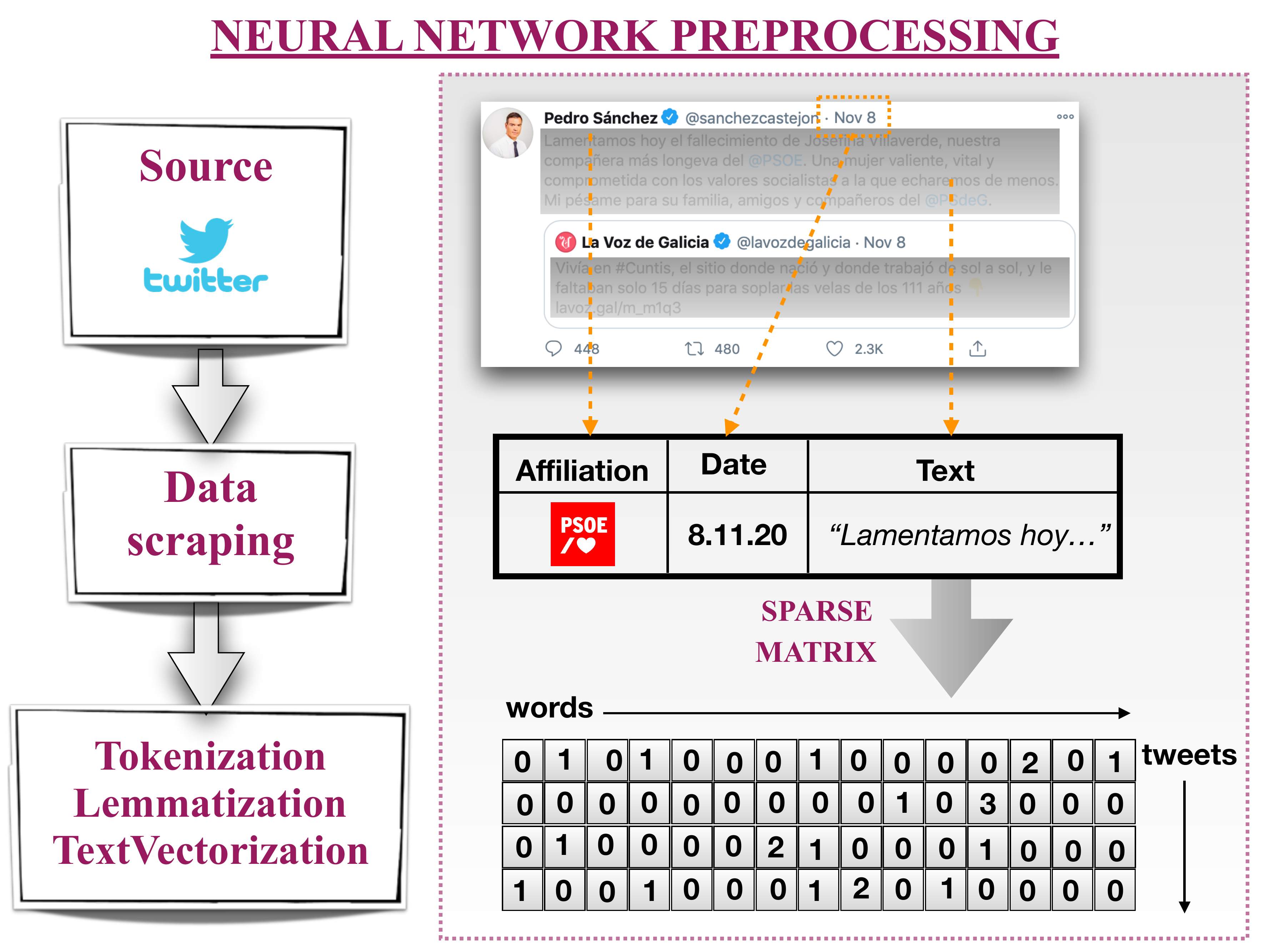}
    \caption{Workflow for data pre-processing. }
    \label{fig:prepro}
\end{figure}
After acquiring the raw data in the form of a collection of tweets, we pre-process as shown schematically in Fig.~\ref{fig:prepro}. For each tweet, we attach a label with the political party it represents and note the date it was written. The contents of the tweet (string of text) are first {\it tokenised}, namely broken down into words, keeping their relative position. For example, the poem line 'So few grains of happiness' becomes a sequence of ('so', 'few', 'grains', 'of', 'happiness'). We clean this sequence further by removing {\it stop words}, e.g. 'of'. Then the items of this sequence are {\it lemmatised}, i.e. some words are substituted by their lemma, e.g. 'happiness' $\rightarrow$ 'happy'. Additionally, we remove short tweets with fewer than seven words.
At the end of this process of data cleaning we were left with about 108,000 tweets for the analysis.

For the neural network predictive analysis described in Sec.~\ref{sec:NNs}, the tweets are substituted by  numerical arrays, with a column for each appearing word, and a value of 0,1,2 \ldots depending on how many times a particular word appears in a tweet. This process is called {\it TextVectorization} and enables a numerical analysis, with the input for the Neural Network made of a set of arrays, encoding each tweet, and a label for the political party.

\section{Ideological bubbles}\label{sec:bubble}

Before embarking into a numerical sentiment analysis and predictive Neural Network algorithms, we can   gain some initial understanding on  possible ideological differences among  parties by using Word Clouds. These clouds are graphical representations of the word frequency and are nowadays quite commonplace. 

\begin{figure}
    \centering
    \includegraphics[width=0.45\textwidth]{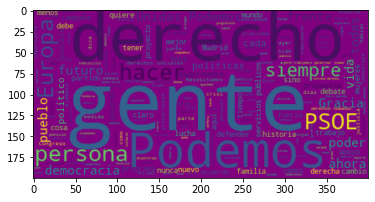}
     \includegraphics[width=0.45\textwidth]{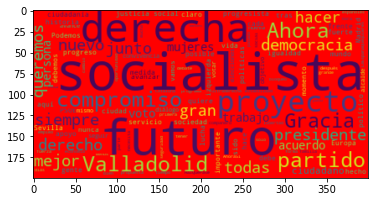}
     \includegraphics[width=0.45\textwidth]{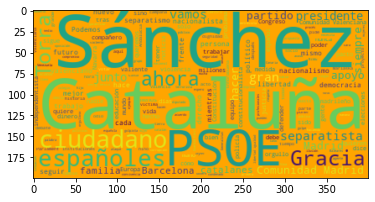}
     \includegraphics[width=0.45\textwidth]{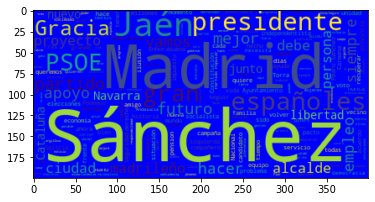}
     \includegraphics[width=0.45\textwidth]{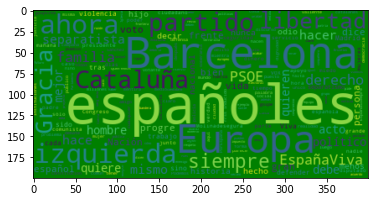}
    \caption{World Clouds for the main political parties in 2019. Top to bottom corresponds to left to right political leaning.}
    \label{fig:clouds}
\end{figure}

In Fig.~\ref{fig:clouds} we show the Word Clouds for the year 2019. 
The higher the recurrence of a particular item in the tweets, the larger the word will appear in the cloud. Parties with very repetitive messages will then show large words, whereas parties with more diverse  messages (vocabulary-wise) will show a cloud with many words. 

The differences among the parties' clouds are obvious.  On the far left of the political spectrum, the world cloud of \Podemos highlights {\it derecho} (legal rights)  , {\it gente} (people) besides its own name. 

\PSOE's word cloud is the most diverse one, although with high recurrence of its own socialist project ({\it proyecto} and {\it socialista}), the opposition's label {\it derecha} (right wing) and the word {\it futuro} (future).

On the centre-right spectrum we find \Ciudadanos and \PP, both quite focused on the governing left party PSOE with words like {\it Sanchez} (PSOE's leader) and {\it PSOE} high in the recurrence list. Nevertheless, there are some differences, with \Ciudadanos quite focused on the Catalonian independence issue ({\it Catalu\~na}), and \PP with {\it Madrid}, where they lead the local government. 

\Vox's cloud also shows high incidence of words related to national identity: {\it Catalu\~na} and {\it Barcelona}, related to the Catalonian independence dispute, as well as the word {\it espa\~noles} (Spaniards) and {\it Europa} (Europe).

For each party, the clouds are very similar each year with the exception of \PSOE's in 2016, which was  the leading opposition party and trying to impeach \PP's leader and president (Mariano Rajoy). The world cloud showed then a high incidence of keywords {\it Rajoy} and {\it Cambio} ('change').

All parties, from far left to far right exhibit very different word clouds, rough representations of their ideological bubbles.

With Data Science tools we can go beyond this naive analysis and quantify deeper issues such as: {\it 1.)} Does the sentiment of a party message change, and is this change a reaction to external events? and {\it 2.)} Is each party's  message distinctive and can be identified tweet-by-tweet?

We will explain methodologies to answer both questions, with the use of Sentiment Analysis and Artificial Intelligence (AI), respectively.

\section{Sentiment Analysis}\label{sec:sentiment}

In this section we explain a procedure to explore the evolution of the sentimental change of   political parties along this last four years. 

Within the area of Data Science, human language is studied under the umbrella of {\it Natural Language Processing} (NLP). With the increasing use of AI, NLP has become a very powerful way to analyse and predict human communication. For example, with NLP tools one can effectively identify fake content~\cite{fake}, predict the next word/sentence in a conversation~\cite{predict}, analyse speech patterns~\cite{speech}, generate new texts~\cite{newtext} among many others. 

For this project we focused on the evolution of {\it sentiment}, namely how positive or negative tweets from party representatives are. Most words do not carry any universal sentimental value, e.g. 'house' is a neutral word for most people, but other words do carry universal sentimental value. For example 'great' and 'horrible' have clear sentiment associations, as well as smiling or vomiting emojis. 

In the area of NLP we have tools  to transform a string of text into a sentiment score, with positive/negative numerical values indicating the level of sentiment. 
For the analysis we will present here, we used a  python library called VADER (Valence Aware Dictionary and sEntiment Reasoner)~\cite{VADER}, which is particularly well adapted to social media contexts. This library is only available in English, hence we first  translate the tweets to English using a python library  {\it translate}~\cite{Gtrans} and perform the analysis on the translated tweets.

\begin{figure}
    \centering
    \includegraphics[scale=0.35]{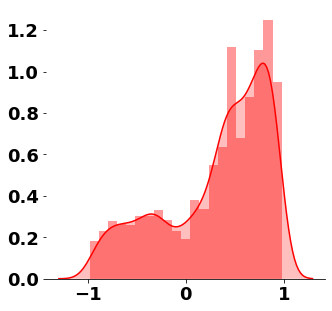}
      \includegraphics[scale=0.35]{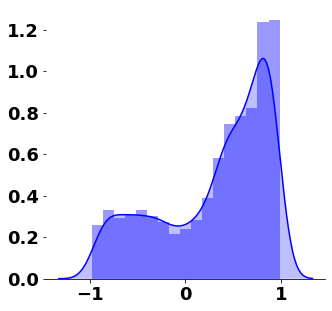}
      \includegraphics[scale=0.23]{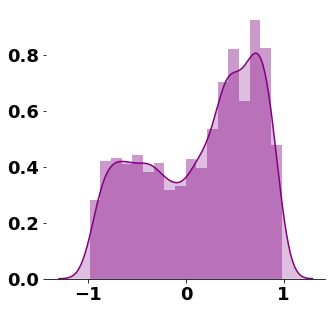}
       \includegraphics[scale=0.23]{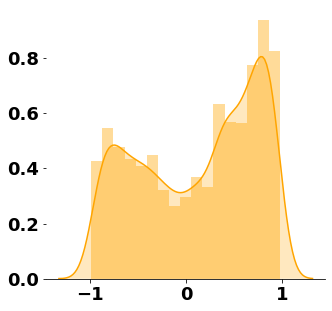}
    \includegraphics[scale=0.23]{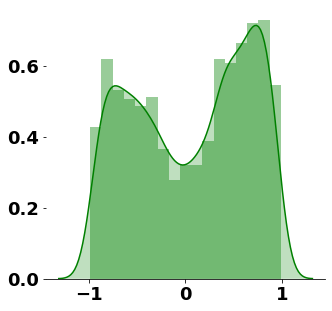}
    \caption{Spanish political parties sentiment distributions during the year 2019. The x-axis correspond to negative and positive sentimental values and the y-axis represents the frequency of this sentiment value. }
    \label{fig:sentimentdistr}
\end{figure}

We can represent sentiment in a number of ways, and here we choose two: {\it 1.)}  distribution of sentiment of the political parties, linked to the level of polarization of their messages, and {\it 2.)} time-evolution of this sentiment as a consequence of external events. 

Firstly, in Fig.~\ref{fig:sentimentdistr} we analyse all tweets from 2019 and plot the sentiment distributions.
The traditional parties, \PSOE and \PP (upper panel), show a clear bias towards positive sentiment. We have observed this behaviour in all years we have analyzed, despite the change in government. On the other hand, newer parties (lower panel) send more polarized messages, where the ratio of negative to positive messages is much higher than for the traditional parties.  In the case of \Podemos this  ratio decreased in 2019, coinciding with their joining of \PSOE in a coalition government. Acquiring governing responsibilities seem to have lowered their level of polarization. 

\begin{figure*}[h!]
    \centering
    \includegraphics[scale=0.45]{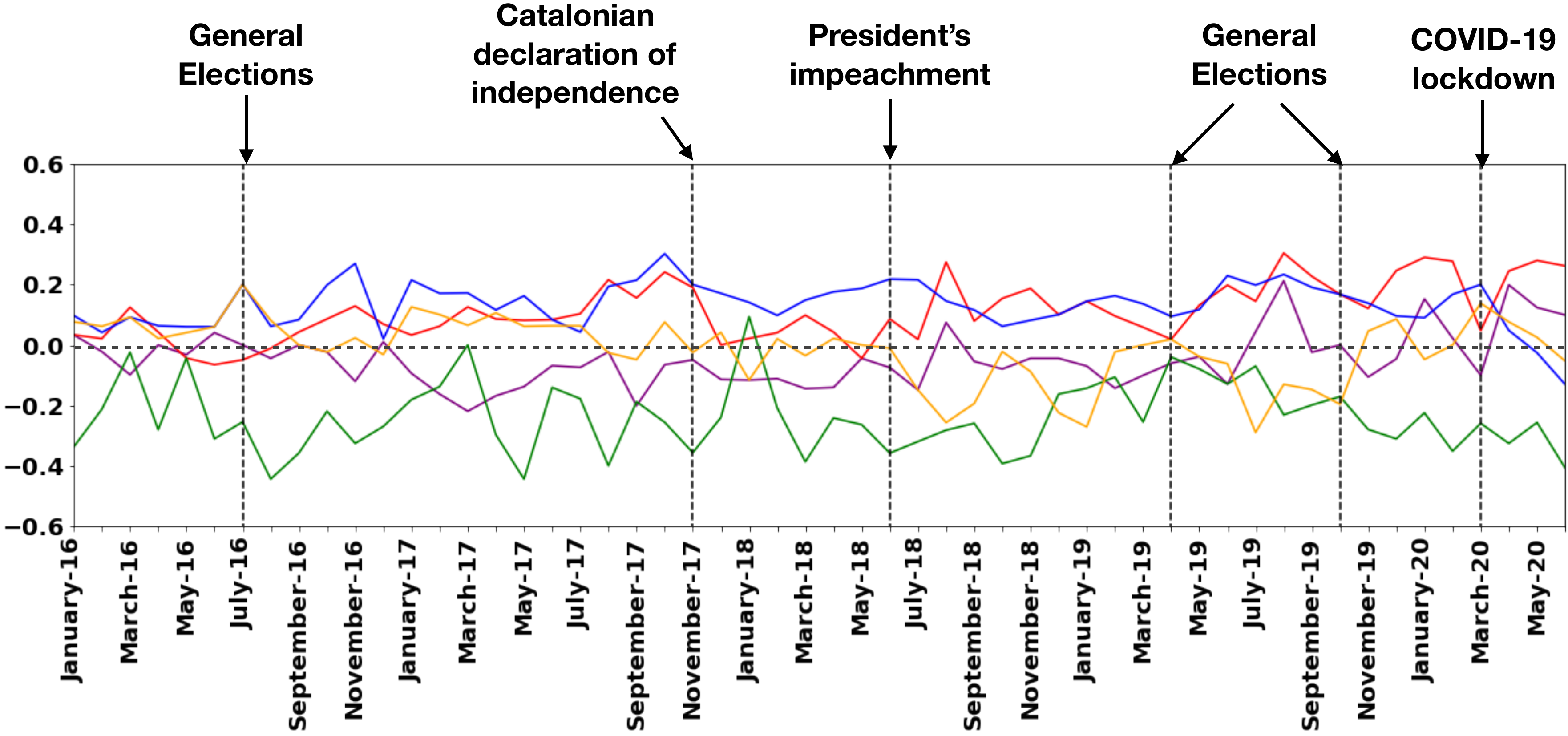}
    \caption{Overall sentiment evolution for each party in the period January 2016 till May 2020. Dashed vertical lines denote particularly important political issues. }
    \label{fig:sentimentevol}
\end{figure*}

This separation of behaviour of well-established and newer parties may be due to the level of political experience in their leadership composition. Established parties may not engage in too-extreme messages to the population, trying to display a position of more balanced and professional communication.

Besides overall positivity and negativity distributions, one can quantitatively investigate whether changes in sentiment could be linked to specific events. To do so, in Fig.~\ref{fig:sentimentevol} we display the monthly median of sentiment level.

There are a number of points to discuss in this figure. In the following discussion, when we talk about a party's positivity or negativity, we simply refer to the value in this curve, it is not meant to be an absolute statement about the party. 

If one pays attention to the horizontal dashed line, corresponding to zero overall sentiment, the red \PSOE and blue \PP lines are above that line for the whole period, except that \PP drops sharply below during the COVID pandemia. The overall positivity of these two traditional parties has already been noted in Fig.~\ref{fig:sentimentdistr}. The jump of \PP to negative values as the lockdown evolved seems to be a reaction to the government's (\PSOE and \Podemos) handling of the pandemia, of which \PP has been extremely critical. 

 New parties are markedly more negative, again a reflection of what we already observed in Fig.~\ref{fig:sentimentdistr}.
  
  Moreover, there are interesting differences among these new parties. \Vox is clearly the most biased towards negative sentiment and it also displays a more variable behaviour. The behaviour of centrist \Ciudadanos was closer to the traditional parties during the period preceding the Catalonian independence proclamation, but at that moment it started  dropping in sentiment values. This negative trend was broken in March 2020, where a new leader was elected for this party. On the other hand, \Podemos shows an interesting evolution: with an overall negative profile, it suddenly starts coupling its behaviour to the other left-wing \PSOE after the right-wing President is successfully impeached. Note how the highs and lows of \Podemos and \PSOE follow the same pattern from that impeachment on.
 
\Vox's sentiment values  after the 2016 election, when this party did not fare particularly well (see Fig.~\ref{fig:evol})  evolved towards more negativity. This negative trend is also apparent during the months preceding the Catalonian declaration of independence. As discussed in the ideological bubble section, territorial identity issues are a strong focus of \Vox. The independence process was squashed, its political leaders were arrested or fled the country. This failure strengthened the ideological position of \Vox, and may be the origin of the unusual level of positivity in the early 2018. Another positive trend appears in the 2019 general elections, when \Vox's representation increased dramatically, see last graph in Fig.~\ref{fig:evol}, becoming the third most voted party in Spain. Following the other opposition parties, \Vox shows a downward trend following the beginning of the COVID pandemic.

Despite \PP's overall positive sentiment, one can observe a negative correlation with \Podemos up to the Catalonian declaration of independence. Up to that point, an upward trend for \PP was paired with a downward evolution for \Podemos and vice-versa. After that point, we see  \PP exhibiting downward trends related to negative events for the party, e.g. the President's impeachment, leading to the loss of the governance for PP. We also see a positive peak in-between the two 2019 general elections (April and November of the same year), when the leading party \PSOE was unable to form a coalition government and re-ran the elections. 

Between the same two 2019 general elections, \Ciudadanos went from 57 representatives to 10. The re-run of the elections was very damaging for this political party but, interestingly, the evolution of the sentiment was not the expected downward trend. 

The sentiment values of\PSOE went up after a successful impeachment, and each general election in 2019 ---when they were the most voted political party. Their sentiment values decreased sharply around the time lockdown was announced by their governing coalition,  and also after Catalonia's independence declaration.

\section{Predictive Neural Network analysis}
\label{sec:NNs}

In the previous section we showed that trends in sentiment values could be matched to specific events, and that some parties' behaviour do correlate and anti-correlate. Particularly interesting is the case of \Podemos, which was anti-correlated with \PP's behaviour until the Catalonian declaration of independence, and correlated with \PSOE from the President Rajoy's impeachment on.  
\begin{figure}[h!]
    \centering
    \includegraphics[width=0.45\textwidth]{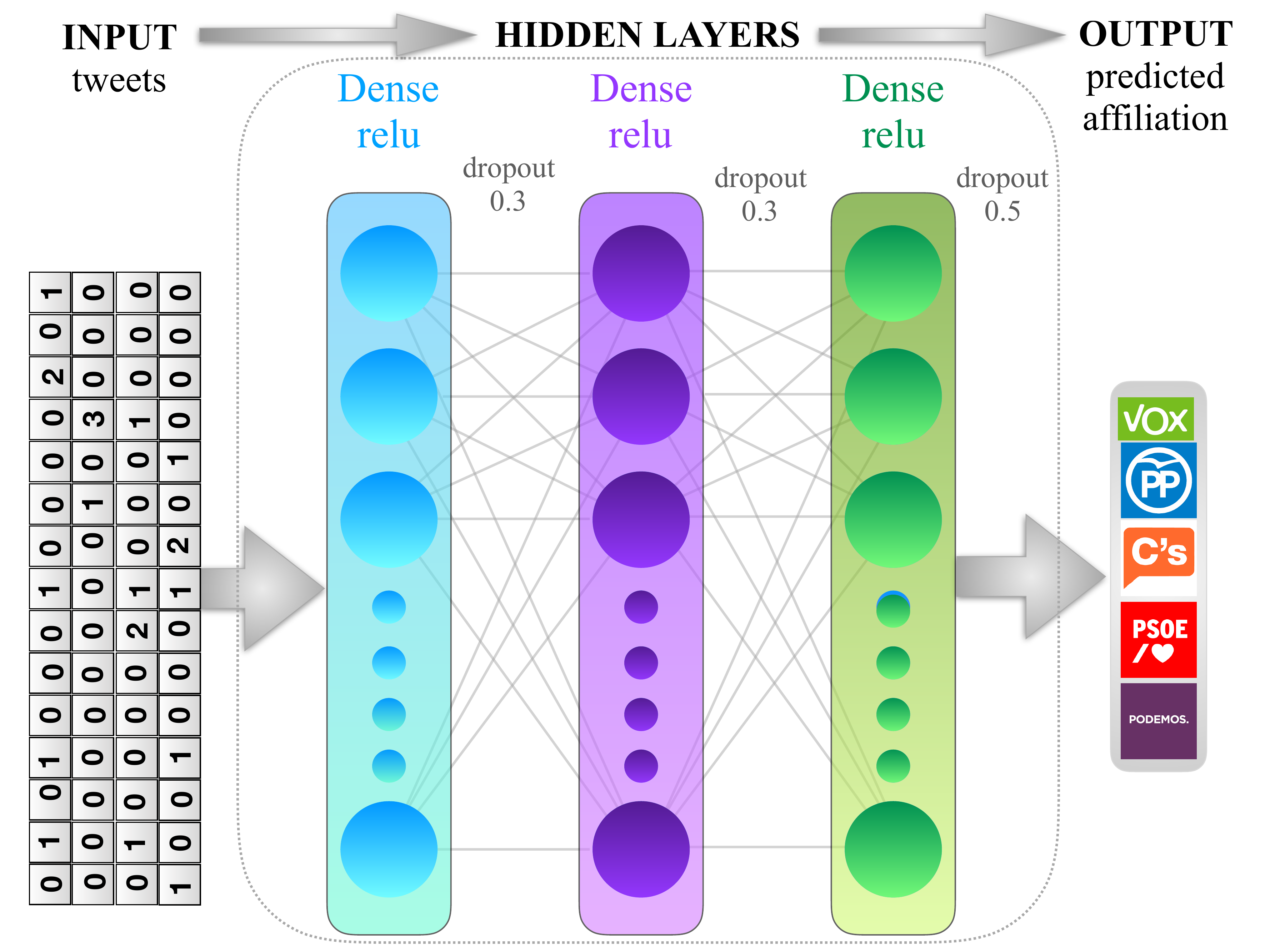}
    \caption{Artificial Intelligence analysis using Neural Networks to perform a supervised classification problem.}
    \label{fig:NNplot}
\end{figure}

In this section we are going to explore another way of looking at the same data. Instead of analysing the sentiment carried by the meaning of words inside tweets, we simply analyse the tweets as strings of word-frequency numbers, as explained in Sec.~\ref{sec:dataselection} and schematically represented in Fig.~\ref{fig:prepro}. With these strings as inputs and the political affiliation as label, we perform a supervised Machine Learning analysis based on Neural Networks, see Fig.~\ref{fig:NNplot}. 

The architecture of our Neural Network consists of an input layer with the vectorised tweets, three hidden layers with 16 neurons each, plus intermediate dropout layers, and a categorical output in the form of one of the following labels: \Vox \PP \Ciudadanos \PSOE \Podemos. This Neural Network is going to learn to classify tweets as belonging to a party. It will train with an amount of data (80\% of the total dataset from the year 2019) and refine an algorithm which provides each individual tweet's probability of belonging to a party. The training in this case is supervised, namely at each iteration of the Neural Network training it gets a feedback on how well it has been able to predict. After many iterations with no increase in accuracy, the values of the algorithm are frozen and used to produce predictions for  new, unseen tweets (20\% of the dataset). To avoid biases in the training, we balance the amount of tweets for each party, in a process called {\it undersampling}.

\begin{figure}[h!]
    \centering
    \includegraphics[width=0.45\textwidth]{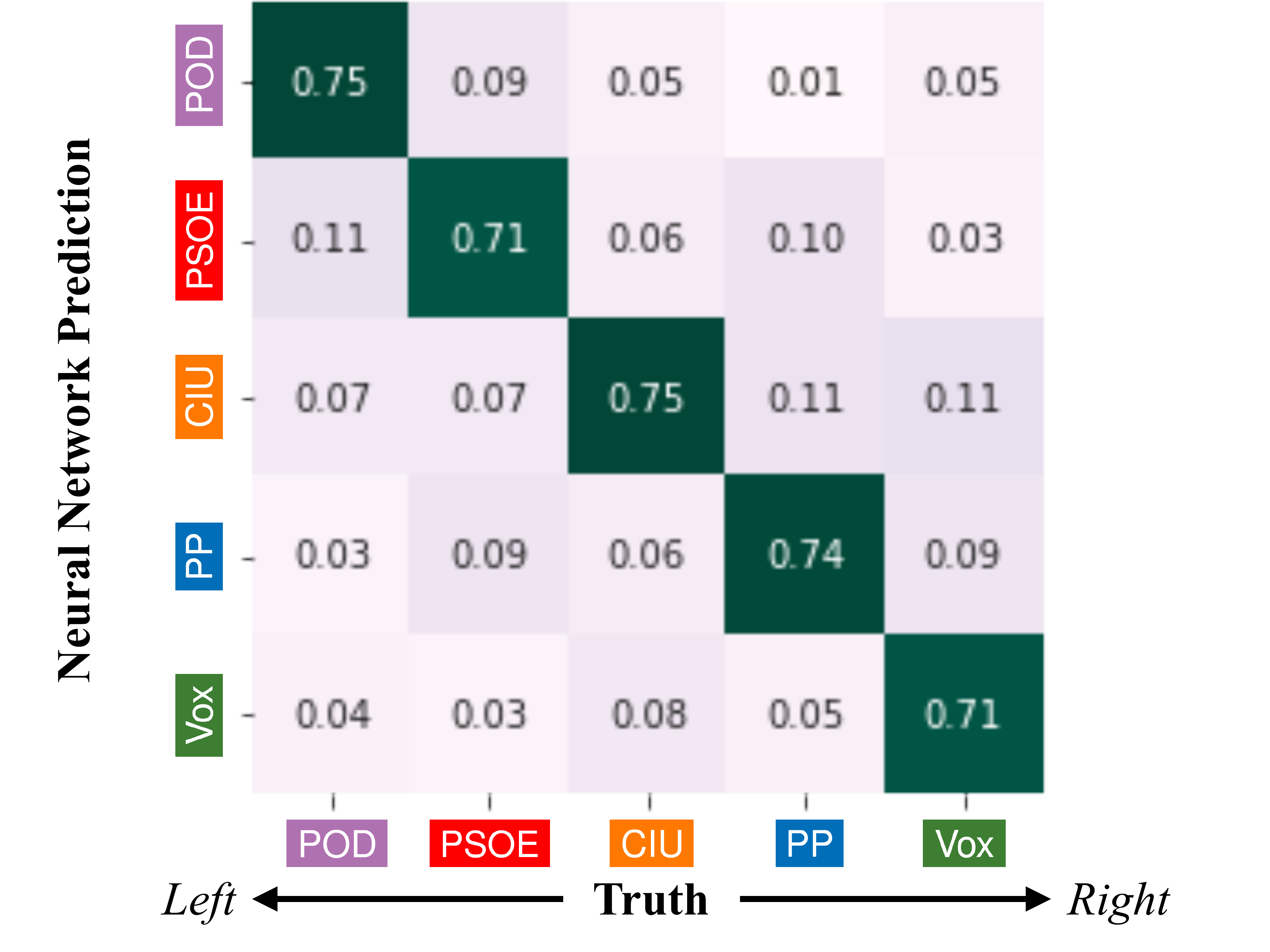}
    \caption{Neural Network predictions: 5$\times$5 confusion matrix.}
    \label{fig:all2019}
\end{figure}

The algorithm's ability to learn can be expressed as a total accuracy, namely how many tweets were correctly assigned. Since we are dealing with a multi-class problem, it is better to express the results in terms of a {\it confusion matrix}, see Fig.~\ref{fig:all2019}. In this figure we show a matrix which contains percentages. The x-axis correspond to true labels, and y-axis is the predicted label. The way to understand this matrix is the following:

The diagonal, deep green, entries show the percentage of tweets that are correctly assigned for each party. For example, a tweet from a leader of \Podemos would be correctly identified 75\% of the time, and one from \Vox 71\%. 

The non-diagonal entries, in light pink colours represent how often tweets from a political party are mis-identified. For example, let us look at the first column, corresponding to \Podemos. These numbers show that a true tweet from this party will be mis-identified three times more often as a tweet from their coalition government partners, \PSOE, than with any of the right-wing parties. 

The algorithm finds a  higher mis-identification  rate of tweets from \Vox to and from \Ciudadanos than with \PP, despite belonging to the far-right and centre respectively. This may be due to their shared strong interest in the Catalonian territorial issues. 

One can find another significant pair of correlations between \PP and \PSOE, which could be traced back to their shared status of traditional party.

\begin{figure}[h!]
    \centering
    \includegraphics[width=0.45\textwidth]{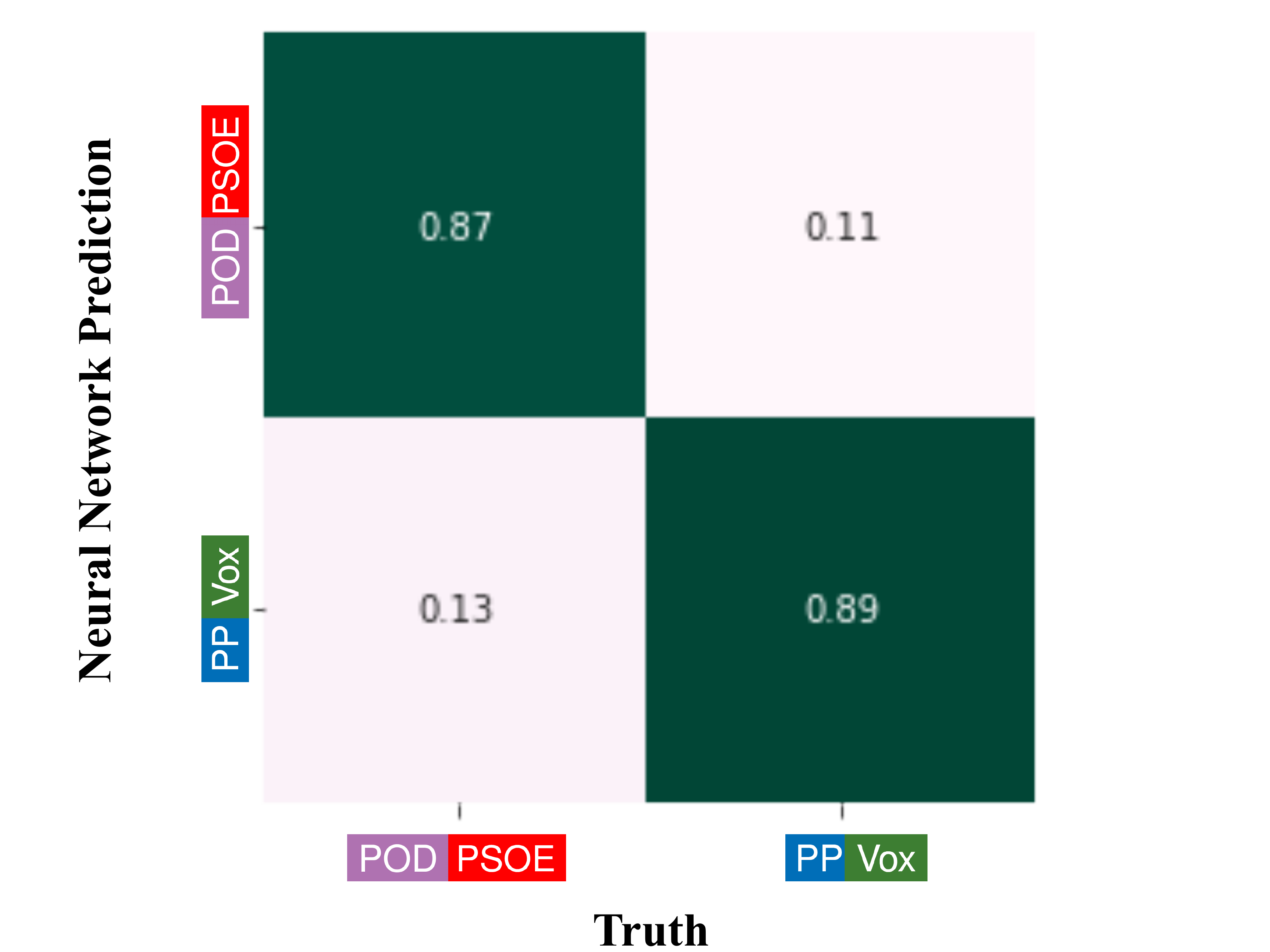}
    \caption{Neural Network predictions: 2$\times$2 confusion matrix.}
    \label{fig:leftright}
\end{figure}

We can also ask the Neural Network a simpler question, whether a tweet is from a left- or right-leaning party. The statistical answer to that question is  provided in Fig.~\ref{fig:leftright}. About 90\% of the time, the algorithm will get the answer right. 

\section{Summary and Outlook}

Data Science is becoming a very powerful tool to investigate human interactions. For example, the development of Natural Language Processing (NLP) techniques allows us to scrape large amounts of language-based interactions and analyse them in ways that a few years back were inconceivable. On top of that, the development of new and more powerful prediction techniques, in particular Neural Networks, opens new possibilities for prediction and search for unknown patterns.

To demonstrate this potentiality, we have chosen a particular set of human interactions, social media outlet via Twitter, and the political arena in our home country, Spain. When trying to check the political pulse, communication by Twitter is subject to lots of noise, but also quite dynamical in response to current events. Data Science is prepared to deal with noisy environments provided enough data is analysed.

We have described several levels of data analysis for the tweets written by Spanish politicians since 2016. We started with a simple analysis of the party ideological bubble by looking at the recurrence of words. From these bubbles we noted a number of salient points: how some parties are more focused on other party's leaders than their own programme, or how some political movements are quite tuned to territorial issues. 

Those bubbles do provide some information, but only on broad features. To go a step further we took two different routes:

{\it 1.)} We analysed levels of positive and negative sentiment in the tweets using new tools adapted to social media. From that study we found a number of very interesting features. We first observed that new and traditional parties formed different sets of sentiment behaviour, with traditional parties more geared towards positive messages. 
But more complex patterns were revealed when we studied the overall party sentiment as a function of time. We observed correlations of this sentiment with political events, as well as in-between parties' behaviour.

{\it 2.)} We analysed the content of the tweets (without assigning any sentimental value) and trained a form of Artificial Intelligence, Neural Network, to recognise the political affiliation of a tweet. The AI was able to predict the origin of the tweet with a precision in the range of 71-75\%, and the political leaning (left or right) with a precision of around 90\%. The high accuracy achieved by the AI implies there is a high level of adherence to specific political views. 
Apart from overall predictions, we also found interesting features on the mis-indentification rates, indicating some level of proximity between parties' agendas, e.g. positions respect to the Catalonian independence claims.

Our analysis has a number of approximations which could be improved upon. Leaders' tweets capture the political pulse of our representatives, but one could expand this study to include the followers of political leaders, and explore to what level there is more general adherence to political views. The sentiment analysis we performed required an automatic translation from Spanish to English which could have added some level of noise, especially when dealing with the informal language present in tweet exchanges. The Neural Network analysis was based on a very simple architecture which we did not tune to achieve the best possible values, and one could expect a few percent improvement by tuning the hyper-parameters. Moreover, in this part of the study, we could have added mixed inputs including other sources beyond tweets.

This study is meant as a proof-of-concept exploration. Our results show that NLP and AI tools can be useful and reliable when dealing with complex human interaction problems. Despite our focus on the Spanish political system in 2016-2019, we believe the range of possible applications is huge. For example, this study is directly translatable to any country with a democratic system and a wide use of Twitter or other social media outlets.

The database and code (in python) to reproduce this analysis will be made public at Miguel Folgado's {\tt GitHub} account ({\tt Miguelfolgado}).


\end{document}